\documentclass[twocolumn,showpacs,preprintnumbers,amsmath,amssymb,superscriptaddress,nofootinbib]{revtex4-1}

\usepackage{graphicx}
\usepackage[british]{babel}

\newcommand{\gsim}{\;\rlap{\lower 3.5 pt \hbox{$\mathchar \sim$}} \raise 1pt
 \hbox {$>$}\;}
\newcommand{\lsim}{\;\rlap{\lower 3.5 pt \hbox{$\mathchar \sim$}} \raise 1pt
 \hbox {$<$}\;}

\newcommand{\lmmOS}{l_{\mbox{\scriptsize OS}}}
\newcommand{\lmmMS}{l_{\mbox{\scriptsize$\overline{\rm MS}$}}}

\usepackage{color}
\newcommand{\num}[1]{\color{black} #1}

\allowdisplaybreaks

\begin{document}


\title{Quark mass relations to four-loop order in perturbative QCD}

\author{Peter Marquard}
\affiliation{Deutsches Elektronen Synchrotron DESY,
  Platanenallee 6, 15738 Zeuthen, Germany}

\author{Alexander V. Smirnov}
\affiliation{Scientific Research Computing Center, Moscow State University,
  119991, Moscow, Russia}

\author{Vladimir A. Smirnov}
\affiliation{Skobeltsyn Institute of Nuclear Physics, Moscow State
  University, 119991, Moscow, Russia}

\author{Matthias Steinhauser}
\affiliation{Institut f{\"u}r Theoretische Teilchenphysik, Karlsruhe
  Institute of Technology (KIT), 76128 Karlsruhe, Germany}

\date{February 2015}

\begin{abstract}
  We present results for the relation between a heavy quark mass defined in
  the on-shell and $\overline{\rm MS}$ scheme to four-loop order.  The method
  to compute the four-loop on-shell integral is briefly described and the new
  results are used to establish relations between various short-distance
  masses and the $\overline{\rm MS}$ quark mass to next-to-next-to-next-to-leading
  order accuracy. These relations play an important role in the accurate
  determination of the $\overline{\rm MS}$ heavy quark masses.
\end{abstract}

\preprint{
  DESY-15-013,
  TTP15-01,
  SFB/CPP-14-120}

 
\maketitle


The precise knowledge of quark masses plays an important role in many
phenomenological applications. This is in particular true for the heavy top,
bottom and charm quarks.  For example, the top quark mass enters as a crucial
parameter the combined electroweak fits which have been used to obtain
indirect information about the Higgs boson mass, and nowadays serve as
consistency checks for the Standard Model, see, e.g.
Refs.~\cite{Schael:2013ita,Ciuchini:2013pca}.
The uncertainty in the top quark mass is also
dominant in the analyses of the stability of the
electroweak vacuum~\cite{Bezrukov:2012sa,Degrassi:2012ry,Alekhin:2012py}. 
A prominent example where a precise bottom quark mass is
required are $B$-meson decays which are often proportional to the fifth power
of $m_b$.  Precise charm and bottom quark masses are important to obtain
accurate predictions for the Higgs boson decays into the respective quark
flavours. Also in the context of top and bottom Yukawa coupling unification,
precise mass values are indispensable since they serve as boundary conditions
at low energies. Last but not least, quark masses enter the Lagrange density
of the Standard Model as fundamental parameters. Thus, it is mandatory to
obtain precise numerical values by comparing high-order theoretical
predictions with precise experimental data.

At lowest order in perturbation theory there is no need to fix the
renormalization scheme for the quark masses. However, after including quantum
corrections it is necessary to apply renormalization conditions which fix the
renormalization scheme. A natural scheme for heavy quark masses, i.e. the
charm, bottom and top quark masses, is the on-shell (OS) scheme
where one requires that the inverse heavy
quark propagator with momentum $q$ has a zero at the position of the on-shell
mass, $M$, i.e. for $q^2=M^2$.  It is well known that perturbation theory has
a bad convergence behaviour in case the on-shell quark mass is used as a
parameter.  Another widely used renormalization scheme is based on minimal
subtraction. This means that the mass parameter entering the quark propagator
is defined in such a way that just divergent terms (and no finite
contributions) are absorbed such that the quark propagator is finite (after
wave function renormalization). In this Letter we consider four-loop
corrections to the relation between the on-shell and the $\overline{\rm MS}$
definition of a heavy quark mass which allows for a precise conversion from one
renormalization scheme into the other.

For the various heavy quarks different methods relying on different quark mass
definitions are used to extract the mass values. For example, in
Ref.~\cite{Chetyrkin:2009fv} low-moment sum rules have been used to extract
directly the $\overline{\rm MS}$ charm and bottom quark masses without any
reference to the on-shell mass. On the other hand, physical observables
inherently connected to the threshold, like $\Upsilon$ sum rules or top quark
pair production close to threshold, rely on properly defined quark masses,
like the potential subtracted (PS)~\cite{Beneke:1998rk},
1S~\cite{Hoang:1998hm,Hoang:1998ng,Hoang:1999zc} or renormalon subtracted
(RS)~\cite{Pineda:2001zq}
definition. When comparing with experimental data, in a first step the
corresponding mass values are extracted.  Afterwards they are converted to the
$\overline{\rm MS}$ definition.  Note that the relation between the
$\overline{\rm MS}$ and the OS mass is an important ingredient to obtain the
relation between the PS, 1S or RS masses and the $\overline{\rm MS}$ mass.

In this Letter we use the four-loop $\overline{\rm MS}$-OS relation to
establish relations between the PS, 1S, RS and the
$\overline{\rm MS}$ quark mass which are necessary to
obtain the latter with next-to-next-to-next-to leading order (N$^3$LO)
accuracy.

Note that there is a further definition of a threshold mass, the
so-called kinetic mass~\cite{Czarnecki:1997sz} which has
been used for quite a number of applications in $B$ physics,
(see, e.g., Ref.~\cite{Gambino:2013rza}).
However, the relation to the on-shell mass is only known to two loops
(NNLO). For this reason it is not considered in the following.

In the following we first discuss the relation between the 
$\overline{\rm MS}$ and OS quark mass. Afterwards we elaborate
on the relation between the threshold (PS, 1S and RS) and the
$\overline{\rm MS}$ mass. The latter is obtained by using as 
starting point the definition of the PS, 1S or RS masses 
which establishes a relation to the pole mass. Afterwards 
the pole mass is replaced by the $\overline{\rm MS}$ mass
which leads to the desired relation between the short-distance
masses. 

To derive a formula relating the $\overline{\rm MS}$ and OS quark mass
it is advantageous to start with relations between theses masses and the bare
mass, $m^0$, which are given by\footnote{We refrain from adding a superscript
  ``OS'' to the on-shell mass but use a capital letter. Similarly, a
  lower-case $m$ without further superscript stands for the $\overline{\rm
    MS}$ quark mass. For all other mass definitions we use a lower-case ``m''
  and a superscript indicating the renormalization scheme.}
\begin{eqnarray}
  m^0 = Z_m^{\overline{\rm MS}}m \,,&\quad&
  m^0 = Z_m^{\rm OS} M\,.
  \label{eq::mbare}
\end{eqnarray}
$Z_m^{\overline{\rm MS}}$ is known to four loops and can be found in
Refs.~\cite{Chetyrkin:1997dh,Vermaseren:1997fq,Chetyrkin:2004mf}. 
By construction, the ratio of the two equations in~(\ref{eq::mbare})
is finite which leads to 
\begin{eqnarray}
   z_m(\mu) &=& \frac{m(\mu)}{M}
   \,,
  \label{eq::OS2MS}
\end{eqnarray}
where $z_m$ depends on $\alpha_s(\mu)$ and $\log(\mu/M)$
and has the following perturbative expansion
\begin{eqnarray}
  z_m(\mu) &=& \sum_{n\ge0} \left(\frac{\alpha_s}{\pi}\right)^n z_m^{(n)}
  \,,
  \label{eq::cm}
\end{eqnarray}
with $z_m^{(0)}=1$.

For a derivation of convenient formulae relating $z_m(\mu)$ to the on-shell
quark self energy we refer to
Refs.~\cite{Gray:1990yh,Melnikov:2000qh,Marquard:2007uj} where it is shown
that $Z_m^{\rm OS}$ is obtained from the sum of scalar and vector contribution
evaluated on-shell, i.e.,
\begin{eqnarray}
  Z_m^{\rm OS} &=& 1 + \Sigma_V(q^2=M^2) + \Sigma_S(q^2=M^2)
  \,.
\end{eqnarray}
One-, two- and three-loop QCD results to $Z_m^{\rm OS}$ have been
computed in Refs.~\cite{Tarrach:1980up},~\cite{Gray:1990yh}
and~\cite{Chetyrkin:1999ys,Chetyrkin:1999qi,Melnikov:2000qh,Marquard:2007uj},
respectively, and electroweak effects have been considered in
Refs.~\cite{Hempfling:1994ar,Jegerlehner:2003py,Faisst:2004gn,Martin:2005ch,Eiras:2005yt}.
The main task of this Letter is the computation of the four-loop QCD
corrections to $Z_m^{\rm OS}$ and consequently to $z_m$.  For convenience we
introduce also the inverse relation to Eq.~(\ref{eq::OS2MS}) as follows
\begin{eqnarray}
  M &=& m(\mu) c_m(\mu)\,.
  \label{eq::MS2OS}
\end{eqnarray}

The PS quark mass has been introduced in Ref.~\cite{Beneke:1998rk}.
Its relation to the pole mass is given by
\begin{eqnarray}
  m^{\rm PS} &=& M - \delta m(\mu_f)
  \,,
  \label{eq::OS2PS}
\end{eqnarray}
with
\begin{eqnarray}
  \delta m(\mu_f) &=& 
  -\frac{1}{2} 
  \int_{|\vec{q}\,|<\mu_f} \frac{{\rm d}^3q}{(2\pi)^3}
  V(\vec{q}\,)
  \,,
\end{eqnarray}
where $V(\vec{q}\,)$ is the perturbative contribution to the static heavy
quark potential. $\delta m(\mu_f)$ can be computed in perturbative QCD
and has the form
\begin{eqnarray}
  \delta m(\mu_f) &=& \mu_f \frac{C_F\alpha_s}{\pi}
  \Bigg\{1 
  + \frac{\alpha_s}{4\pi} 
  \left[a_1 + \beta_0 \left( 2 + \log\frac{\mu^2}{\mu_f^2} \right)
  \right]
  \nonumber\\&&\mbox{}
  + \ldots
  \Bigg\}
  \,,
\end{eqnarray}
where $\beta_0=11- 2n_l/3$ is the one-loop coefficient of the 
QCD $\beta$ function and 
$a_1= 31/3 - 10 n_l/9$
the one-loop coefficient of the static potential.
$n_l$ is the number of massless quarks.
$\mu_f$ is the factorization scale which is of the order of the soft scale.
In this Letter we use $\mu_f=2$~GeV for bottom and $\mu_f=20$~GeV
for top quarks.  $\delta m(\mu_f)$ is known to N$^3$LO~\cite{Beneke:2005hg}
which involves the three-loop corrections to the static potential,
$a_3$~\cite{Smirnov:2008pn,Smirnov:2009fh,Anzai:2009tm}.

The N$^3$LO relation between $m^{\rm PS}$ and $m$ is obtained by inserting
Eq.~(\ref{eq::MS2OS}) into Eq.~(\ref{eq::OS2PS}).  All ingredients are already
expanded and the coefficients of $\left(\alpha_s\right)^n$ have simply to be
combined. In particular, the term in~(\ref{eq::OS2PS}) involving $a_3$ is
combined with the four-loop term in the $m$-$M$ relation. One obtains an
explicit formula to compute the PS mass in case the 
$\overline{\rm MS}$ is given. For a given PS mass we solve this
equation iteratively to obtain the $\overline{\rm MS}$ mass.

The 1S mass is defined as half the perturbative mass of a fictitious $1^3S_1$
state, where it is assumed that the quark is stable. Thus, we have the
following relation between the 1S and on-shell
mass~\cite{Hoang:1998hm,Hoang:1998ng,Hoang:1999zc}
\begin{eqnarray}
  m^{\rm 1S} &=& M 
  + \frac{1}{2} 
  E_1^{\rm pt}\Big|_{\alpha_s^n \to \alpha_s^n \varepsilon^{n-1}} 
  \,,
  \label{eq::OS21S}
\end{eqnarray}
where $E_1^{\rm pt}$ is the perturbative ground state energy which is
available to third order~\cite{Penin:2002zv,Beneke:2005hg,Kiyo:2014uca}.  The
last missing ingredient was the three-loop static potential which has been
evaluated in Refs.~\cite{Smirnov:2008pn,Smirnov:2009fh,Anzai:2009tm}.  The
replacement $\alpha_s^n \to \alpha_s^n \varepsilon^{n-1}$ implements the
so-called $\varepsilon$ expansion which guarantees that the appropriate orders
in the expansions of $E_1^{\rm pt}$ and $M$ (in terms of $m$) are combined.

The perturbative expansion of $E_1^{\rm pt}$ has the following form
\begin{eqnarray}
  E_1^{\rm pt}\Big|_{\alpha_s^n \to \alpha_s^n \varepsilon^{n-1}}
  &=&
  - \varepsilon \frac{C_F^2 M\alpha_s^2}{8} 
  \sum_{n\ge0}  \left(\varepsilon \frac{\alpha_s}{\pi}\right)^n \delta_E^{(n)}
  \,.
\end{eqnarray}

In order to obtain the relation between $m^{\rm 1S}$ and $m$ 
one has to replace $\alpha_s^{(n_l+1)}$ in Eq.~(\ref{eq::MS2OS}) by
$\alpha_s^{(n_l)}$ and then apply the replacement $\alpha_s^n \to \alpha_s^n
\varepsilon^{n}$. Afterwards it is inserted into Eq.~(\ref{eq::OS21S}) and
expanded in the parameter $\varepsilon$.  This guarantees that the N$^k$LO
term in $E_1^{\rm pt}$ is combined with the $(k+1)$-loop correction to
$c_m(\mu)$. In particular, in order to establish the $m^{\rm 1S}$-$m$ relation
to N$^3$LO four-loop corrections to $c_m(\mu)$ are needed.  A given
$\overline{\rm MS}$ quark mass is transformed to the 1S mass by inserting the
numerical value into the resulting equation.  In case the 1S mass is given one
obtains the $\overline{\rm MS}$ mass by solving the equation implicitly.

A further threshold mass, the so-called RS mass, has been
introduced in Ref.~\cite{Pineda:2001zq}. It is related to the pole mass in
such a way that the pure renormalon contributions are subtracted.  The
corresponding formulae are derived and explicitly given in
Ref.~\cite{Pineda:2001zq}. In that reference also a variant, the so-called
RS$^\prime$ scheme, is discussed where no subtraction is performed for the
${\cal O}(\alpha_s)$ term in the $\overline{\rm MS}$-OS relation.
Recently the numerical accuracy of the normalization constant of the first
renormalon has been improved in Ref.~\cite{Ayala:2014yxa}, where also a
variant of the RS and RS$^\prime$ masses has been suggested in which (in the
case of the bottom quark) the subtraction term is parameterized in terms of
$\alpha_s^{(3)}$.  In this paper we will adopt the prescription of
Ref.~\cite{Pineda:2001zq}.  Similarly to the PS mass also for the RS mass a
subtraction scale has to be specified which we again choose as $\mu_f=2$~GeV
for bottom and $\mu_f=20$~GeV for top.

For the computation of the scalar and vector part of the fermion propagator we
use an automated setup which generates all contributing amplitudes, processes
them with {\tt FORM3}~\cite{Vermaseren:2000nd} and provides scalar functions
involving several million different integrals encoded in functions with 
14 different indices which belong to 100 different integral families. 

The Laporta algorithm~\cite{Laporta:2001dd} is applied to each family using
{\tt FIRE5}~\cite{Smirnov:2014hma} and {\tt crusher}~\cite{crusher} 
which are written in {\tt C++}.  Then we use
the code {\tt tsort}~\cite{Pak:2011xt}, which is part of the latest {\tt FIRE}
version, to reveal relations between primary master integrals following
recipes of~\cite{Smirnov:2013dia} and end up with 386 four-loop massive
on-shell propagator integrals, i.e. with $p^2=M^2$.

We have performed the calculation allowing for a general gauge
parameter $\xi$ keeping terms up to order $\xi^2$ in the expression we
give to the reduction routines. We have checked that $\xi$ drops out
after mass renormalization but before inserting the master integrals.

For some master integrals, analytic results could be derived using a
straightforward loop-by-loop integration for general space-time dimension.  We
also used analytical results obtained for non-trivial four-loop on-shell
master integrals computed in our earlier paper Ref.~\cite{Lee:2013sx}.  In
some other cases one- and two-fold Mellin-Barnes representations can be
derived which allow for a high-precision numeric evaluation, at least up to 20
digits.  For some of the master integrals, we applied threefold MB
representations which enabled us to obtain a precision of eight digits.

For factorizable integrals, we obtained analytic results from
known two- and three-loop results. In particular, we used 
Ref.~\cite{Lee:2011jt} where the expansion in $\epsilon=(4-d)/2$ has been 
performed up to the order typical to four-loop calculations.
($d$ is the space-time dimension used to compute the momentum integrals.)

We computed the remaining 332 integrals numerically with the
help of {\tt
  FIESTA}~\cite{Smirnov:2008py,Smirnov:2009pb,Smirnov:2013eza}.  {\tt
  FIESTA} returns for each $\epsilon$ coefficient a numerical result
and the corresponding uncertainty from the numerical integration. When
inserting the master integrals we keep track of all uncertainties and
combine them quadratically in the final expression.  We interpret the
resulting uncertainty as a standard deviation and multiply it by five
in the final result for the relation between the
$\overline{\rm MS}$ and OS quark mass.
This is in agreement with adding the uncertainties from the individual
contributions linearly.

We are now in the position to present numerical results for $z_m(\mu)$ which
have been obtained by setting the number of colours to three ($N_c=3$) and the
number of massless quarks ($n_l$) to either 3, 4 or 5, corresponding to the
charm, bottom or top quark case, before combining the uncertainties from the
numerical integration of the master integrals.  Note that the coefficients up
to three loops are known analytically~\cite{Melnikov:2000qh,Marquard:2007uj}.
We refrain from listing the corresponding results but refer to Eq.~(13) of
Ref.~\cite{Chetyrkin:2000yt}.  Analytical results are also available for the
logarithmic four-loop contributions since they can easily be obtained using
renormalization group methods.  In the following we restrict ourselves to
compact numerical results. At four loops we obtain for the coefficient of
$(\alpha_s/\pi)^4$ 
\begin{eqnarray}
  z_m^{(4)}\Big|_{n_l=3} &=& -1744.8 \pm 21.5
  -   703.48\,\lmmOS
  -   122.97\,\lmmOS^2
  \nonumber\\&&\mbox{}
  -   14.234\,\lmmOS^3 
  -   0.75043 \,\lmmOS^4
  \,,\nonumber\\
  z_m^{(4)}\Big|_{n_l=4} &=& -1267.0 \pm 21.5
  - 500.23  \,\lmmOS
  - 83.390  \,\lmmOS^2
  \nonumber\\&&\mbox{}
  - 9.9563 \,\lmmOS^3 
  - 0.514033 \,\lmmOS^4
  \,,\nonumber\\
  z_m^{(4)}\Big|_{n_l=5} &=& -859.96 \pm 21.5 
  - 328.94  \,\lmmOS
  - 50.856  \,\lmmOS^2
  \nonumber\\&&\mbox{}
  - 6.4922  \,\lmmOS^3 
  - 0.33203 \,\lmmOS^4
  \,,
  \label{eq::zm4}
\end{eqnarray}
with $\lmmOS = \ln(\mu^2/M^2)$. We obtain the $\mu$-independent coefficients
with an accuracy of {\num 1.2\%} for $n_l=3$, {\num 1.7\%} for $n_l=4$) and 
{\num 2.5\%} for $n_l=5$.
In the numerical results discussed below we will 
assume a relative uncertainty of {\num 3\%} for all values of $n_l$.

For convenience we also show the four-loop results for $c_m$
which read
\begin{eqnarray}
  c_m^{(4)}\Big|_{n_l=3} &=& 1691.2 \pm 21.5
  +  828.43 \,\lmmMS
  +  189.65 \,\lmmMS^2
  \nonumber\\&&\mbox{}
  +  36.688 \,\lmmMS^3 
  +  4.8124 \,\lmmMS^4
  \,,\nonumber\\
  c_m^{(4)}\Big|_{n_l=4} &=& 1224.0 \pm 21.5
  + 601.98 \,\lmmMS
  + 134.10 \,\lmmMS^2
  \nonumber\\&&\mbox{}
  + 28.846 \,\lmmMS^3 
  + 3.9648 \,\lmmMS^4
  \,,\nonumber\\
  c_m^{(4)}\Big|_{n_l=5} &=& 827.37 \pm 21.5
  + 408.88 \,\lmmMS
  + 86.574 \,\lmmMS^2
  \nonumber\\&&\mbox{}
  + 22.023 \,\lmmMS^3 
  + 3.2227 \,\lmmMS^4
  \,,
  \label{eq::cm4}
\end{eqnarray}
with $\lmmMS = \ln(\mu^2/m^2)$.
In the remaining part of this Letter we will concentrate on the top and bottom
quark mass. 

As an application of the new results in Eqs.~(\ref{eq::zm4}) and~(\ref{eq::cm4})
we study the relations between the various threshold masses and the 
$\overline{\rm MS}$ mass.
We use the following input values for the strong coupling constant and the
bottom and top quark masses:
\begin{eqnarray}
  \alpha_s^{(5)}(M_Z) &=& 0.1185~\mbox{\cite{Agashe:2014kda}}\,,\quad
  m_b(m_b) = 4.163~\mbox{GeV~\cite{Chetyrkin:2009fv}}\,,\nonumber\\
  M_t &=& 173.34~\mbox{GeV~\cite{ATLAS:2014wva}}\,.
\end{eqnarray}
$\alpha_s$ with four and six active flavours is obtained from
$\alpha_s^{(5)}$ where for the decoupling scale we choose twice
the heavy quark mass~\cite{Chetyrkin:2000yt,Schmidt:2012az}.

Let us have a closer look to the relation between the OS and
$\overline{\rm MS}$ top quark mass. For $\mu=m_t$ we have
\begin{eqnarray}
  M_t &=& m_t\left(
    1 + 0.4244 \,{\alpha_s} + 0.8345 \,{\alpha_s^2} + 2.375 \,{\alpha_s^3}
  \right.\nonumber\\&&\mbox{}\left.
    + (8.49 \pm 0.25) \,{\alpha_s^4}
  \right)
  \nonumber\\&=&\mbox{}
  163.643 + 7.557 + 1.617 + 0.501
  \nonumber\\&&\mbox{}
  + 0.195 \pm 0.005~\mbox{GeV}
  \,,
  \label{eq::mt}
\end{eqnarray}
with ${\alpha_s}\equiv \alpha_s^{(6)}(m_t) = {\num 0.1088}$.
Note that the four-loop term still gives a contribution of about
{\num 200~MeV} which is not negligible even with nowadays uncertainties from
TEVATRON and LHC~\cite{ATLAS:2014wva}.
The corresponding results for the bottom quark read
\begin{eqnarray}
  M_b &=& m_b\left(
    1 + 0.4244 \,{\alpha_s} + 0.9401 \,{\alpha_s^2} + 3.045 \,{\alpha_s^3}
  \right.\nonumber\\&&\mbox{}\left.
    + (12.57 \pm 0.38) \,{\alpha_s^4}
  \right)
  \nonumber\\&=&\mbox{}
  4.163 + 0.401 + 0.201 + 0.148
  \nonumber\\&&\mbox{}
  + 0.138 \pm 0.004~\mbox{GeV}
  \,.
  \label{eq::mb}
\end{eqnarray}
Here ${\alpha_s}\equiv \alpha_s^{(5)}(m_b)={\num 0.2268}$.  Note that the
four-loop corrections in Eq.~(\ref{eq::mb}) are almost as large as the
three-loop term. On the other hand, the perturbative series for the case of
the top quark has a reasonable behaviour: the three-loop coefficient is by a
factor three smaller than the two-loop one and the four-loop term is again
smaller by a factor 2.5.  This suggests that with the help of
Eq.~(\ref{eq::mt}) the top quark mass can be determined with an uncertainty
below 200~MeV.

In practice it often happens that in a first step a threshold quark mass is
extracted from comparisons of higher order calculations and experimental
measurements. Afterwards the threshold mass is converted to the $\overline{\rm
  MS}$ quark mass. In Tabs.~\ref{tab::mtMS} and~\ref{tab::mbMS} we show the
results for the scale invariant $\overline{\rm MS}$ quark mass $m_q(m_q)$
($q=b,t$) using one- to four-loop accuracy for the conversion.

\begin{table}[ht]
\begin{center}
\begin{tabular}{c|ccc}
input & $m^{\rm PS} =$ & $m^{\rm 1S} = $ & $m^{\rm RS} =$ \\
\#loops  &    171.792 &    172.227 &    171.215\\
\hline
1       &    165.097&    165.045&    164.847 \\
2       &    163.943&    163.861&    163.853 \\
3       &    163.687&    163.651&    163.663 \\
4       &    163.643&    163.643&    163.643 \\
\hline
4 ($\times 1.03$)       &    163.637&    163.637&    163.637 \\
\end{tabular}
\caption{\label{tab::mtMS}$m_t(m_t)$ in GeV computed from the PS, 1S and RS quark
  mass using one- to four-loop accuracy. The numbers in the last line 
  are obtained by increasing the four-loop coefficient  in Eq.~(\ref{eq::cm4})
  by {\num 3\%}.}
\end{center}
\end{table}

In the case of the top quark (cf. Tab.~\ref{tab::mtMS}) the three-loop
corrections amount to about {\num 200-250~MeV} which reduces to {\num
  $\{44,8,20\}$~MeV} at four loops for the $\{$PS,1S,RS$\}$ quark mass. A
{\num 3\%} uncertainty in the $\overline{\rm MS}$-OS relation induces a shift
of {\num 6~MeV} in $m_t(m_t)$ which is in general small as
compared to the four-loop contribution. Let us estimate the final uncertainty
from the conversion to the $\overline{\rm MS}$ mass from the quadratic
combination of the {\num 6~MeV} with half of the four-loop contribution
(i.e. {\num $\{44,8,20\}{\times 1/2}$~MeV}). This leads to {\num
  $\{23,7,11\}$~MeV} which should be added in quadrature to the remaining
uncertainties of the threshold mass.

\begin{table}[ht]
\begin{center}
\begin{tabular}{c|ccc}
input & $m^{\rm PS} =$ & $m^{\rm 1S} = $ & $m^{\rm RS} =$ \\
\#loops  &      4.483 &      4.670 &      4.365\\
\hline
1       &      4.266    &      4.308    &      4.210 \\
2       &      4.191    &      4.190    &      4.172 \\
3       &      4.161    &      4.154    &      4.158 \\
4       &      4.163    &      4.163    &      4.163 \\
\hline
4 ($\times 1.03$)       &      4.159    &      4.159    &      4.159 \\
\end{tabular}
\caption{\label{tab::mbMS}$m_b(m_b)$ in GeV computed from the PS, 1S and RS quark
  mass using one- to four-loop accuracy. The numbers in the last line 
  are obtained by increasing the four-loop coefficient in Eq.~(\ref{eq::cm4}) 
  by 3\%.}
\end{center}
\end{table}

The results for $m_b(m_b)$ computed from the PS, 1S and RS threshold masses
are shown in Tab.~\ref{tab::mbMS}. The three-loop corrections provide still
sizable effects of up to {\num 40~MeV} which reduces to at most {\num 9~MeV}
at four loops.  The uncertainty in the four-loop $\overline{\rm MS}$-OS
relation induces an error of {\num 4~MeV}.  Thus we arrive at a final error of
{\num $\{4,6,5\}$~MeV} for the conversion from the $\{$1S,PS,RS$\}$ mass.
This is not negligible, though in general much smaller than other
uncertainties involved in the quark mass extraction (see, e.g.,
Refs.~\cite{Hoang:2012us},~\cite{Ayala:2014yxa} and~\cite{Beneke:2014pta} for
recent determinations of $m_b(m_b)$ where in intermediate steps the 1S, RS and
PS has been used, respectively).

The results of Tabs.~\ref{tab::mtMS} and~\ref{tab::mbMS} can be used, in
combination with similar calculations for different values of $\alpha_s(M_Z)$
and threshold masses, to construct the following approximation formulae
\begin{eqnarray}
  \frac{m_t(m_t)}{\mbox{GeV}} &=& 163.643 \pm 0.023 + 0.074 \Delta_{\alpha_s}
  - 0.095 \Delta_{m_t}^{\rm PS}
  \,,\nonumber\\
  \frac{m_t(m_t)}{\mbox{GeV}} &=& 163.643 \pm 0.007 + 0.069 \Delta_{\alpha_s}
  - 0.096 \Delta_{m_t}^{\rm 1S}
  \,,\nonumber\\
  \frac{m_t(m_t)}{\mbox{GeV}} &=& 163.643 \pm 0.011 + 0.067 \Delta_{\alpha_s}
  - 0.095 \Delta_{m_t}^{\rm RS}
  \,,\nonumber\\
  \frac{m_b(m_b)}{\mbox{GeV}} &=& 4.163 \pm 0.004 + 0.007 \Delta_{\alpha_s}
  - 0.018 \Delta_{m_b}^{\rm PS}
  \,,\nonumber\\
  \frac{m_b(m_b)}{\mbox{GeV}} &=& 4.163 \pm 0.006 + 0.008 \Delta_{\alpha_s}
  - 0.019 \Delta_{m_b}^{\rm 1S}
  \,,\nonumber\\
  \frac{m_b(m_b)}{\mbox{GeV}} &=& 4.163 \pm 0.005 + 0.004 \Delta_{\alpha_s}
  - 0.018 \Delta_{m_b}^{\rm RS}
  \label{eq::m_par}
\end{eqnarray}
with {\num
$\Delta_{\alpha_s} = (0.1185-\alpha_s(M_Z))/0.001$,
$\Delta_{m_t}^{\rm PS} = (171.792~\mbox{GeV} - m_t^{\rm PS})/0.1$,
$\Delta_{m_t}^{\rm 1S} = (172.227~\mbox{GeV} - m_t^{\rm 1S})/0.1$,
$\Delta_{m_t}^{\rm RS} = (171.215~\mbox{GeV} - m_t^{\rm RS})/0.1$,
$\Delta_{m_b}^{\rm PS} = (4.483~\mbox{GeV} - m_b^{\rm PS})/0.02$,
$\Delta_{m_b}^{\rm 1S} = (4.670~\mbox{GeV} - m_b^{\rm 1S})/0.02$,
$\Delta_{m_b}^{\rm RS} = (4.365~\mbox{GeV} - m_b^{\rm RS})/0.02$. }

\begin{table}[ht]
\begin{center}
  \begin{tabular}{c||l|l|l}
    Ref. & \multicolumn{3}{c}{$c_m^{(4)}(\mu=m(m))$} \\
    \hline 
    & $n_l=3$ & $n_l=4$ & $n_l=5$ \\
    \hline 
    \cite{Beneke:1994qe} & 1668 & 1324 & 1031 \\
    \cite{Chetyrkin:1997wm} & $1571.4$ & $1107.8$ & $727.0$  \\
    \cite{Kataev:2010zh} & 1281 & 986 & 719  \\
    \cite{Ayala:2012yw} & 1785.9 & 1316.4 & 920.1 \\
    \cite{Sumino:2013qqa} & $1668 \pm 167$ & $1258^{+26}_{-66}$ & $897^{+31}_{-175}$ \\
    \cite{Ayala:2014yxa} & $1772 \pm 82$ & $1324 \pm 82$ & $945 \pm 92$ \\
    \hline
    this work & $1691.2 \pm 21.5$ & $1224.0 \pm 21.5$ & $827.37 \pm 21.5$ \\ 
  \end{tabular}
  \caption{\label{tab::compare}Comparison of predictions for the four-loop
    coefficient $c_m^{(4)}$ to our result which is present in the last line.}
\end{center}
\end{table}

Let us finally compare in Tab.~\ref{tab::compare} our result for the four-loop
coefficient $c_m^{(4)}$ to predictions obtained on the basis of different
assumptions.  In general good agreement is found, in particular with the
results from Refs.~\cite{Ayala:2012yw,Sumino:2013qqa,Ayala:2014yxa} which are
all based on renormalon cancellation. For example, in
Ref.~\cite{Sumino:2013qqa}, the four-loop coefficient is extracted from the
requirement of perturbative stability of the combination $2m_{\rm pole}+V_{\rm
  QCD}$ where $V_{\rm QCD}$ is the static potential of two heavy quarks.
The estimates in Ref.~\cite{Chetyrkin:1997wm} have been obtained
only on the basis of the two-loop results, leading nevertheless to good
approxima\-tions.
Somewhat lower results have been obtained in Ref.~\cite{Kataev:2010zh} where
dispersive methods have been used and large $\pi^2$ terms have been
identified.  In Ref.~\cite{Beneke:1994qe} the four-loop relation between the
on-shell and $\overline{\rm MS}$ quark mass has been estimated using the large
$\beta_0$ approximation.

To conclude, in this Letter we have computed the four-loop corrections
between the on-shell and $\overline{\rm MS}$ definition of heavy quarks.
Our main results are given in Eqs.~(\ref{eq::zm4}) and~(\ref{eq::cm4})
for charm, bottom and top quarks. As applications we have derived precise
relations between the PS, 1S and RS threshold masses and the $\overline{\rm
  MS}$ quark mass. 


\section*{Acknowledgments}

We are thankful to Tobias Huber for advice on numerical evaluation of
Mellin-Barnes integrals. M.S. would like to thank Konstantin Chetyrkin,
Hans K\"uhn and Jan Piclum for useful discussions.  
We thank Konstantin Chetyrkin and Kirill Melnikov for useful
comments to the manuscript.
We thank the High Perfomance Computing Center
Stuttgart (HLRS) and the Supercomputing Center of Lomonosov Moscow
State University~\cite{LMSU} for providing computing time used for the
numerical computations with {\tt FIESTA}.  P.M was supported in part
by the EU Network HIGGSTOOLS PITN-GA-2012-316704.  This work was
supported by the DFG through the SFB/TR~9 ``Computational Particle
Physics''.  The work of V.S. was supported by the Alexander von
Humboldt Foundation (Humboldt Forschungspreis).



\end{document}